\documentclass{aastex63}

\usepackage{amsmath}
\usepackage{times}
\usepackage{mathptmx}
\usepackage[caption=false]{subfig}
\DeclareMathAlphabet{\mathcal}{OMS}{cmsy}{m}{n}
\DeclareSymbolFont{largesymbols}{OMX}{cmex}{m}{n}

\received{\today}
\revised{\today}
\accepted{XXX}

\shorttitle{A search for radio pulses from GRBs with MWA}
\shortauthors{Xu et al.}

\begin{document}

\title{A Search for Low-frequency Radio Pulses from Long Gamma-ray Bursts with the Murchison Widefield Array }

\correspondingauthor{Fan Xu (carpedieminreality@163.com)}

\author[0000-0001-7943-4685]{Fan Xu$^\star$}
\affiliation{International Centre for Radio Astronomy Research, Curtin University, GPO Box U1987, Perth, WA 6845, Australia}
\affiliation{School of Astronomy and Space Science, Nanjing University, Nanjing 210023, People's Republic of China}
\affiliation{Department of Physics, Anhui Normal University, Wuhu, Anhui 241002, People's Republic of China}

\author[0000-0001-6544-8007]{G. E. Anderson}
\affiliation{International Centre for Radio Astronomy Research, Curtin University, GPO Box U1987, Perth, WA 6845, Australia}

\author{Jun Tian}
\affiliation{International Centre for Radio Astronomy Research, Curtin University, GPO Box U1987, Perth, WA 6845, Australia}
\affiliation{Jodrell Bank Centre for Astrophysics, Department of Physics and Astronomy, The University of Manchester, Manchester M13 9PL, UK}

\author{B. W. Meyers}
\affiliation{International Centre for Radio Astronomy Research, Curtin University, GPO Box U1987, Perth, WA 6845, Australia}

\author{S. J. Tingay}
\affiliation{International Centre for Radio Astronomy Research, Curtin University, GPO Box U1987, Perth, WA 6845, Australia}

\author{Yong-Feng Huang}
\affiliation{School of Astronomy and Space Science, Nanjing University, Nanjing 210023, People's Republic of China}
\affiliation{Key Laboratory of Modern Astronomy and Astrophysics (Nanjing University), Ministry of Education, People's Republic of China}

\author{Zi-Teng Wang}
\affiliation{International Centre for Radio Astronomy Research, Curtin University, GPO Box U1987, Perth, WA 6845, Australia}

\author{B. Venville}
\affiliation{International Centre for Radio Astronomy Research, Curtin University, GPO Box U1987, Perth, WA 6845, Australia}

\author{C. P. Lee}
\affiliation{International Centre for Radio Astronomy Research, Curtin University, GPO Box U1987, Perth, WA 6845, Australia}

\author{A. Rowlinson}
\affiliation{Anton Pannekoek Institute for Astronomy, University of Amsterdam, Science Park 904, NL-1098 XH Amsterdam, the Netherlands}
\affiliation{ASTRON, Netherlands Institute for Radio Astronomy, Postbus 2, 7990 AA, Dwingeloo, The Netherlands}

\author{P. Hancock}
\affiliation{International Centre for Radio Astronomy Research, Curtin University, GPO Box U1987, Perth, WA 6845, Australia}
\affiliation{Curtin Institute for Data Science, Curtin University, GPO Box U1987, Perth, WA 6845, Australia}

\author{A. Williams}
\affiliation{International Centre for Radio Astronomy Research, Curtin University, GPO Box U1987, Perth, WA 6845, Australia}

\author{M. Sokolowski}
\affiliation{International Centre for Radio Astronomy Research, Curtin University, GPO Box U1987, Perth, WA 6845, Australia}

\begin{abstract}

It has been proposed that coherent radio emission could be emitted during or shortly following a gamma-ray burst (GRB). 
Here we present a low-frequency ($170-200$ MHz) search for radio pulses associated with long-duration GRBs using the Murchison Widefield Array (MWA). The MWA, with its rapid-response system, is capable of performing GRB follow-up observations within approximately $30$ seconds. Our single pulse search, with temporal and spectral resolutions of $100\ \mu$s and $10$ kHz, covers dispersion measures up to $5000$ pc cm$^{-3}$. Two single pulse candidates are identified with significance greater than $6\sigma$, surviving a friends-of-friends analysis. We rule out random fluctuations as their origin at a confidence level of $97\%$ ($2.2\sigma$). We caution that radio frequency interference from digital TV (DTV) is most likely the origin of these pulses since the DTV frequency bands almost cover the entire observing frequency band. If they are astrophysical signals, we estimate the peak flux densities for our pulse candidates of $3.6\pm0.6$ Jy and $10.5\pm1.5$ Jy, with corresponding fluences of $431\pm74$ Jy ms and $211\pm37$ Jy ms, respectively. Based on these observations and the assumption of a magnetar origin for the pulse, we constrain the radio emission efficiency as $\epsilon_{\rm{r}}\sim10^{-3}$ for both candidates, which is consistent with pulsar observations. Our results highlight the promising potential of new-generation radio telescopes like the MWA to probe the central engines of GRBs.

\end{abstract}


\section{Introduction} \label{sec:intro}

Gamma-ray bursts (GRBs) are extremely energetic events in the universe. Generally, GRBs are classified into two categories, i.e., the long GRB (LGRB), which is spectrally soft with a duration longer than $2$\,s and the short GRB (SGRB), which is spectrally hard with a duration less than $2$\,s \citep{Kouveliotou..1993}.
It is believed that the majority of LGRBs come from massive stellar explosions \citep{woosley93,Galama..1998,Iwamoto..1998,kulkarni98,woosley06heger,Woosley..2006}, while SGRBs are thought to originate from compact stellar mergers \citep{lattimer76,eichler89,narayan92,mochkovitch93,Gehrels..2005,Abbott..2017}. However, these classifications have been challenged by recent observations of SGRBs with extended emission \citep{Norris..2006,Gehrels..2006,Rastinejad..2022,Yang..2022,Yang..2024,Levan..2024} and LGRBs that lack supernovae association \citep{Fynbo..2006,Della..2006,Michalowski..2018}. Whether these bursts should be classified as distinct categories of GRBs is still under debate. Although the prompt radiation of GRBs is primarily in gamma rays, the multi-wavelength afterglow, especially in X-rays, optical, and radio, has provided us with rich information about GRB environments and their progenitors \citep[e.g.][]{wijers97,waxman98,galama99}. Usually, radio afterglows last longer than optical and X-ray afterglows \citep{Frail..1997} and are, therefore, widely used in the study of the late-time blast wave evolution from relativistic to sub-relativistic speed \citep[e.g.][]{Frail..2004,vanderhorst08}. Typically, radio afterglows for LGRBs peak at $3-6$ days at $8.5$ GHz \citep{Chandra..2012,AndersonGE..2018}, consistent with the prediction from the forward external shock \citep{Sari..1998,Zhang..2018}. On the other hand, some GRBs show evidence of early radio flares, approximately $1$ day post-burst \citep{Kulkarni..1999,frail00RS,berger03,AndersonGE..2014,anderson24}. These early radio flares are commonly explained by a reverse external shock \citep{Sari..1999}.

It has been long proposed that bright short-duration radio emission could be detected associated with GRBs at earlier times, i.e., prompt emission \citep{Baird..1975}. The signal may be very similar to fast radio bursts (FRBs) with millisecond duration and several models predict such emissions \citep[see][for a detailed review]{Rowlinson..2019}, with a possible detection from the short GRB 201006A \citep{rowlinson24}. 
Despite the denser burst environment of LGRBs, which makes such emissions harder to escape compared to SGRBs, the higher event rate of LGRBs increases the chance of detection. Therefore, this study focuses on LGRBs. There are three main phases during which the coherent radio emission from LGRBs may occur. First, after the relativistic jet has been launched, coherent radio emission could be produced from the interaction between the jet and interstellar medium (ISM) \citep{Usov..2000,Sagiv..2002}. Second, bright pulsar-like radio emission may be emitted from a long-lived millisecond magnetar central engine \citep{Totani..2013}. A long-lived millisecond magnetar central engine can naturally explain the plateaus \citep{Zhang..2006,Rowlinson..2010,Rowlinson..2013} and X-ray flares \citep{Burrows..2005,Falcone..2007,Yi..2022} observed in many GRB afterglows \citep{Dai..1998,Zhang..2001,Rowlinson..2014,Starling..2020,Xu..2021}. Third, coherent radio emission could occur when the rapidly spinning-down magnetar fails to support itself and collapses into a black hole \citep{Falcke..2014,Zhang..2014}.

Previous efforts to detect these bright short-duration radio emissions from LGRBs have yielded negative results. Early searches usually lack sensitivity or perform follow-up observations too slowly to capture the prompt emission expected to be emitted seconds to minutes after the initial gamma-ray flash \citep{Koranyi..1995,Dessenne..1996,Benz..1998,Bannister..2012,Palaniswamy..2014}. More recent searches using new-generation radio telescopes are capable of performing rapid-response observations and have much higher sensitivity than before \citep{Obenberger..2014,Kaplan..2015,Rowlinson..2019b,Anderson..2021pasa,Rowlinson..2021,Tian..2022pasa,Tian..2022mn,Curtin..2023,Hennessy..2023}. Radio emissions experience delays with decreasing frequency due to dispersion, causing signals from GRBs to arrive several seconds to minutes after initial alerts at low frequencies. Therefore low frequency search (below $200$ MHz) would allow us to be on target before the pulsed signal arrives. One of the ideal radio telescopes to search for low-frequency radio pulses from GRBs is the Murchison Widefield Array (MWA; \cite{Tingay..2013,Wayth..2018}). The MWA is capable of beginning observations within $30$ seconds of receiving a GRB alert \citep{Hancock..2019,Anderson..2021pasa,Tian..2022pasa,Tian..2022mn}. This rapid response, combined with the dispersion delay of extra-galactic signals at low frequencies, guarantees that the telescope is on target before the signals arrive. It provides observational data with high temporal resolution ($100$ $\rm{\mu s}$) and frequency resolution ($10$ kHz) \citep{Tremblay..2015}. These features enhance the MWA's sensitivity to detect millisecond-duration transient signals associated with GRBs, allowing it to capture signals emitted within seconds post-burst.

Here we present low-frequency radio pulse searches of
a sample of six LGRBs detected by the Neil Gehrels \emph{Swift} Observatory (hereafter \emph{Swift}; \cite{Gehrels..2004}) with the MWA. In Section \ref{sec:obs}, we describe the observations of our GRB sample and data processing pipeline. The search results are shown in Section \ref{sec:res}. Then, in Section \ref{sec:mod}, we derive parameter constraints by comparing our observation results with theoretical models. We discuss our results in Section \ref{sec:discuss}. The conclusion and future prospects are given in Section \ref{sec:conclu}.

\section{Observations and Analysis} \label{sec:obs}

\subsection{MWA rapid response observations}

The MWA is a low-frequency ($80-300$ MHz) precursor to the Square Kilometre Array (SKA; \cite{Dewdney..2009}). It was intended primarily as an imaging telescope with a low temporal resolution of $0.5$ s \citep{Tingay..2013}. A Voltage Capture System (VCS) mode enhances the MWA's sensitivity for millisecond-duration transient signals due to high temporal and frequency resolutions ($100$ $\rm{\mu s}/10$ kHz) \citep{Tremblay..2015}. Furthermore, with the upgrade of the rapid-response system \citep{Hancock..2019}, the MWA has been able to automatically trigger Virtual Observatory Events (VOEvents; \cite{Seaman..2011}). Normally, it takes tens of seconds to receive a \emph{Swift} VOEvent broadcast. Then, within $30$ s, the MWA will point to the source and start observations \citep{Hancock..2019,Anderson..2021pasa,Tian..2022pasa,Tian..2022mn}. Taking advantage of the rapid-response system and the VCS mode, the MWA becomes an ideal radio telescope to search for low-frequency radio pulses from GRBs.

Our GRB sample contains six well-localized LGRBs detected by the \emph{Swift} Burst Alert Telescope (BAT; \cite{Barthelmy..2005}) during the period from $2022$ to $2023$. 
The localization presented in Table \ref{tab:1} are obtained from the \emph{Swift} X-ray Telescope (XRT, \cite{Burrows..2005b}), enhanced using the \emph{Swift} Ultra-violet Optical Telescope (UVOT, \cite{Roming..2005}) where possible\footnote{\url{https://www.swift.ac.uk/xrt\_positions/}} (see \cite{Goad..2007,Evans..2009}). Exceptions to this are GRB $220427$A and GRB $220518$A, for which the \emph{Swift}-UVOT positions of the optical counterparts are used (\cite{GCN..31960} and \cite{GCN..32063}, respectively).
We also list the detailed prompt emission observational information reported by \emph{Swift} in the General Coordinate Network (GCN) Circulars\footnote{\url{https://gcn.gsfc.nasa.gov/circulars}}. 
All triggered MWA observations were conducted at a central frequency of $185$ MHz with a $30.72$ MHz bandwidth in the MWA phase II extended configuration \citep{Wayth..2018} in VCS mode. 
For each GRB trigger, the observation lasted for $15$ minutes. The delay time between the \emph{Swift} GRB detection and the MWA observation start time for each LGRB in our sample is listed in Table \ref{tab:2}, most of which are $\sim1$\,minute. In Table \ref{tab:2}, we also list the delay time between GCN alert time and the MWA observation start time. All of our observations start $\sim10-30$\,s after receiving the alert, which is sufficient to enable searches for early-time, low-frequency coherent radio emissions from LGRBs.
Based on the information provided by GCN Circulars, none of these bursts have a redshift measurement.

\begin{table}[h!]
	\renewcommand{\thetable}{\arabic{table}}
	\centering
	\tabcolsep=5pt
	\renewcommand\arraystretch{1.2}
	\caption{\emph{Swift} observation information of LGRBs in our sample. } \label{tab:1}
	\begin{tabular}{lccccccc}
		\hline
		\hline
		GRB name & BAT $T_{90}^{a}$ & RA$^{b}$ & Dec$^{b}$ & Uncertainty$^{b}$ & BAT fluence ($15-150$ keV) & BAT photon index & Ref.$^{c}$ \\
		&  (s) & (deg) & (deg) & ($^{\prime}$$^{\prime}$) & ($10^{-7}$ erg cm$^{-2}$) & &  \\
		\hline
		GRB 220427A &  57.2  $\pm$ 12.2  & 275.88882 & -56.25507 & 0.61 & 21   $\pm$ 2    & 1.23  $\pm$ 0.13  & 1, 2 \\
		GRB 220518A &  12.29 $\pm$  3.07 & 55.34289 & -47.58342 & 0.55 &  5.7 $\pm$ 0.8  & 1.71  $\pm$ 0.21  & 3, 4 \\
		GRB 221028A &  57.1  $\pm$ 15    & 314.2026 &  41.0897 & 3.5 & 15.1 $\pm$ 1.4  & 1.367 $\pm$ 0.325 & 5 \\
		GRB 230123A &   5.37 $\pm$  1.03 & 40.85766 & -65.68617 & 1.9 &  2.7 $\pm$ 0.4  & 2.06  $\pm$ 0.24  & 6 \\
		GRB 230204A & 182.05 $\pm$ 16.73 & 238.19965 & -50.91673 & 1.4 & 37   $\pm$ 3    & 1.72  $\pm$ 0.13  & 7 \\
		GRB 230216A &  91.2  $\pm$  7.31 & 113.95902 &  -8.01154 & 1.5 & 13   $\pm$ 2    & 1.9   $\pm$ 0.23  & 8 \\
		\hline
	\end{tabular}	
	\begin{flushleft}
		\tablenotetext{a}{The time interval in the band of $15-150$ keV where 5\%-95\% of the event's fluence was contained (LGRBs are considered to have $T_{90}>2$ s).}
		\tablenotetext{b}{The localization uncertainty represents the radius of the 90\% error region. }	
		\tablenotetext{c}{References for the burst information:(1) \cite{GCN..31960}; (2) \cite{GCN..31968}; (3) \cite{GCN..32063}; (4) \cite{GCN..32069}; (5) \cite{GCN..32893}; (6) \cite{GCN..33214}; (7) \cite{GCN..33279}; (8) \cite{GCN..33336}.  }	  
	\end{flushleft}	
\end{table}

\begin{table}[h!]
	\renewcommand{\thetable}{\arabic{table}}
	\centering
	\tabcolsep=2pt
	\renewcommand\arraystretch{1.2}
	\caption{MWA observation information of LGRBs in our sample. } \label{tab:2}
	\begin{tabular}{lcccccc}
		\hline
		\hline
		GRB name  & ObsID & Full delay$^{a}$ (s) & MWA re-pointing delay$^{b}$ (s) & Sky temperature (K) & Cal$^{c}$ & Used tile number\ /\ Total tile number$^{d}$ \\
		\hline
		GRB 220427A & 1335128544 & 92 & 8  & 444.4 & HerA &  77 / 128 \\
		GRB 220518A & 1336890016 & 37 & 11 & 146.4 & PicA &  72 / 136 \\
		GRB 221028A & 1350998248 & 43 & 22 & 238.5 & 3C444 & 124 / 136 \\
		GRB 230123A & 1358504456 & 54 & 35 & 167.8 & PicA & 114 / 128 \\
		GRB 230204A & 1359514728 & 93 & 19 & 680.0 & HerA & 128 / 144 \\
		GRB 230216A & 1360594192 & 60 & 27 & 206.0 & HydA & 130 / 144 \\
		\hline
	\end{tabular}	
	\begin{flushleft}
		\tablenotetext{a}{The time interval between the \emph{Swift} GRB detection time and the MWA observation start time for each LGRB.}
		\tablenotetext{b}{The time interval between the GCN alert time and the MWA observation start time for each LGRB.}		
		\tablenotetext{c}{The calibrators used to calibrate the MWA observations. }		
		\tablenotetext{d}{The number of good tiles that we used to produce the science-ready data and the total number of tiles used in the observation.}	
	\end{flushleft}
\end{table}



\subsection{VCS data processing} \label{sec:vcs}

We perform the calibration and beamforming of the VCS data following the processing method outlined in \cite{Tian..2022mn}. The MWA VCS data processing has a well-developed pipeline that was initially developed for pulsar science \citep[e.g.,][]{Bhat..2016,McSweeney..2017,Meyers..2017,Ord..2019}. 
We calibrated our VCS data with {\sc mwa\_hyperdrive}\footnote{\url{https://github.com/MWATelescope/mwa\_hyperdrive}}, which is the latest generation calibration software. It provides direction-independent calibration solutions based on a sky model containing bright calibrator sources. The name of the calibrator source used to calibrate each observation is listed in Table \ref{tab:2}. We inspected the solutions for each tile and flagged those with bad solutions that failed to converge. 
Using the delay model and the complex gain information obtained from the solution we then performed coherent beamforming. 
The 
coherent beam width is approximately $1$ arcmin in the extended array configuration of the MWA \citep{Swainston..2022}, which is sufficient to cover the localization region based on the \emph{Swift} or multi-wavelength follow-up positional uncertainties. The final output is time-frequency intensity data with a $100\ \rm{\mu}$s temporal resolution and $10$ kHz frequency resolution. The next step is to generate dedispersion trials and search for single pulses.

\subsection{Single-pulse search} \label{sec:search}

The single-pulse search is conducted using the {\sc PRESTO} software package  \citep{Ransom..2001}. The initial step is to dedisperse the time-frequency data to many trial dispersion measure (DM) values. Due to the lack of redshift measurements for our GRBs, a broad DM search range is set. The lower DM boundary includes contributions from the Milky Way and its halo. For each GRB, the lower DM boundary is different and depends on its position in the sky. We use {\sc PyGEDM} \citep{Price..2021} to calculate the Galactic contribution considering the NE$2001$ model \citep{Cordes..2002} and the halo contribution considering the YT$2020$ model \citep{Yamasaki..2020}. The minimum lower DM boundary for our sample is $100$ pc cm$^{-3}$. We set the upper DM boundary considering the typical range of redshifts for LGRBs. According to \cite{Lan..2021}, $95\%$ of \emph{Swift} LGRBs are within a maximum redshift of $z\sim5$. Using the scaling relation between the DM contribution from the Intergalactic medium (IGM) and redshift as $\rm{DM_{IGM}}\sim 960z$ pc cm$^{-3}$ \citep{Inoue..2004,Batten..2021}, we get a DM upper boundary of $5000$ pc cm$^{-3}$. This DM value corresponds to a delay time of about $10$ minutes between our observing frequency ($185$ MHz) and the gamma-ray frequency, allowing us to detect signals emitted at the time of the explosion within our 15-minute observation window. The DM searching range for each GRB is listed in Table \ref{tab:3}. We use {\sc DDplan.py} from {\sc PRESTO} to decide the search trials, each of which requires different downsampling factors and DM values for downsampling and dedispersion. A dedispersion plan is generated by {\sc DDplan.py}, providing detailed information about the DM step sizes and downsampling factors for each step. With this dedispersion plan, we use {\sc prepsubband} from {\sc PRESTO} to downsample and dedisperse the time-frequency data. As a result, the time-frequency data will be converted into the dedispersed time series data.

The next step is to search for single pulses within the dedispersed time series. This is accomplished with {\sc single\_pulse\_search.py} from {\sc PRESTO} which conducts a sliding boxcar convolution search of time series data with different width boxcar functions. 
Following the procedure outlined in previous single pulse searches \citep{Deneva..2009,Burke-Spolaor..2011,Bannister..2012},
we used the `friends-of-friends' algorithm {\sc rrattrap.py} from {\sc PRESTO}
that identifies when three or more events above $6\sigma$ occur nearly simultaneously with similar DMs, which we then define as a candidate. 
For candidates passing the simple clustering method above, a ``refined'' analysis is performed to enhance the recovered signal-to-noise ratio (SNR) and obtain more precise DMs. We refined the search with a much narrower DM range, smaller DM steps, and more boxcar function options. In the refined analysis, the DM search range is around $2$ pc cm$^{-3}$, with the central DM obtained from the results of clustering method and the DM step is around $0.01$ pc cm$^{-3}$.

To evaluate the chances of a dedispersed time series generating a candidate pulse from statistical variations, we reverse each time series and perform the same pulse search procedure described above, including the use of {\sc rrattrap.py} to identify adjacent events. In other words, we dedisperse the time series with (unphysical) negative DMs \citep{Tian..2022mn,Tian..2022pasa}. Any reversed time series that returns events passing the {\sc rrattrap.py} test implies that the pulse candidates from that observation are more suspect and less likely to be real \citep{Tian..2022pasa}.

\subsection{RFI mitigation} \label{sec:rfi}

\begin{figure}[htbp]
	\centering
	\includegraphics[width=0.8\linewidth]{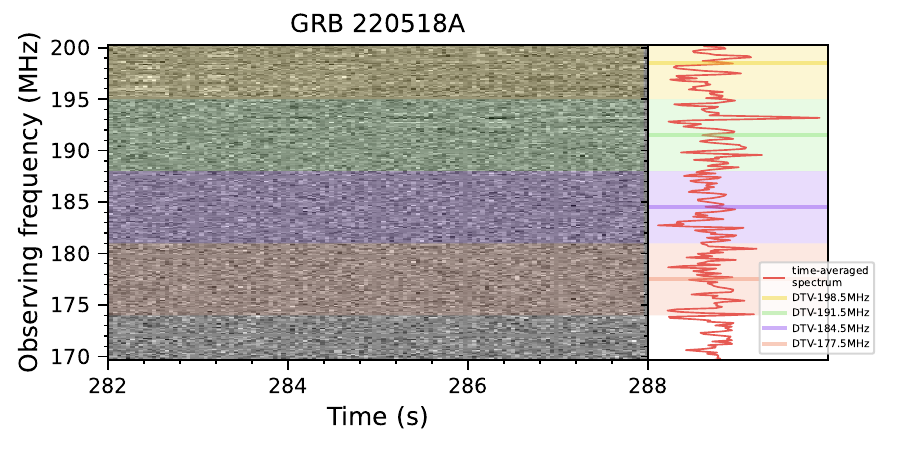}
	\caption{Example of RFI at $\sim 193$ MHz. We used {\sc rfifind} from {\sc PRESTO} to identify and eliminate such RFI. The possibly relevant DTV frequency ($191.5$ MHz) is shown with a green region. The other three DTV emitting bands with central frequencies of $177.5$ MHz, $184.5$ MHz, and $198.5$ MHz are represented by pink, purple, and yellow regions, respectively. On the right side of the panel, we show the time-averaged spectrum and the DTV frequency bands. } \label{fig:1}
\end{figure}

Pulsed broad-band RFI usually appears as an over-dispersed signal after dedispersion, making it straightforward to differentiate from an astrophysical signal. On the other hand, several narrow-band impulsive RFIs may align after dedispersion, creating a false detection. The {\sc rfifind} from {\sc PRESTO} specifically targets periodic RFI. While it can identify and eliminate some impulsive RFI (as shown in Figure \ref{fig:1}), it cannot remove all of them. Figure \ref{fig:1} shows an example of narrow band impulsive RFI occurring at about $284$ s and $286$ s at $\sim 193$ MHz found in the zero DM sequence data of GRB $220518$A with {\sc rfifind}. Additionally, the {\sc rfifind} also generates a mask file to flag this RFI.

The MWA is generally less affected by RFI when compared to other radio telescopes due to its remote location in a designated RFI quiet zone. We therefore performed RFI removal with caution to avoid losing weak signals. We used {\sc rfifind} to eliminate RFI in the zero DM sequence data, where the RFI is at the highest level. 
We acknowledge this as a limitation of our analysis, as there is no easily applicable algorithm for excising impulsive RFI in time series data.

The frequency range of our observations is $170-200$ MHz, within which there are known sources of 
digital TV (DTV) RFI. 
Four DTV emitting bands 
fall within our observing range with central frequencies of $177.5$ MHz, $184.5$ MHz, $191.5$ MHz, $198.5$ MHz and bandwidths of $\sim7$ MHz \citep{Offringa..2015} \footnote{\url{https://ozdigitaltv.com/transmitters/WA}}. The narrow band RFI shown in Figure \ref{fig:1} (within the shaded green region) is probably attributed to DTV reflections from meteor trails \citep{Kemp..2024}. 
Given the large overlap between the DTV emitting bands and our observing band, all pulses discovered in our analysis have to be carefully scrutinized as RFI contamination (see Section \ref{sec:res} for further discussion). 

\subsection{Determination of system sensitivity} \label{sec:sen}
 
We converted the SNR derived from our single pulse search to a flux density 
at the central frequency of our observation ($185$ MHz) using the radiometer equation following the same procedure outlined in \citet[][see equations 1 and 2]{Tian..2022mn}. For our observing frequency, we derived the overall system equivalent flux density by taking the 
antenna efficiency from \cite{Ung..2019},
a typical tied-array gain of $G \sim 0.225$ K Jy$^{-1}$ \citep{Meyers..2017}, an ambient temperature of $T_{\rm{amb}}=300 \pm 15$ K, and receiver temperature of $T_{\rm{rec}}=23$ K. 
To precisely derive the antenna temperature, the standard approach involves calculating the convolution of the antenna pattern with the sky temperature ($T_{\rm{sky}}$) \citep[see][for more details]{Meyers..2017}. While this convolution method provides an accurate calculation, it is computationally expensive. In this study, we instead employ an approximation, where the antenna temperature is estimated as $T_{\rm{ant}}=T_{\rm{sky}} \pm 0.2 T_{\rm{sky}}$, with the corresponding $T_{\rm{sky}}$ used 
for each GRB observation listed in Table \ref{tab:2}. 

For LGRBs with pulse candidates, we calculated the peak flux density $S_{\nu}$ using their SNR from the refined search (see Section \ref{sec:search}). The integration time is taken as 
$100\ \rm{\mu}$s $\times$ downsampling factor. The effective bandwidth of our observation is $26.88$ MHz after flagging edge channels. 
As only a fraction of good tiles were used in our analysis, we adjusted for the corresponding sensitivity loss
by applying a scaling factor equal to the ratio between the number of tiles used and the total number of MWA tiles. 
Note that several tiles were impacted by lightning and shut down during the observations of GRB $220427$A and GRB $220518$A, which resulted in a 40\% loss in sensitivity.
For the other GRBs, the sensitivity remained around $90$\%. After the correction, we arrive at an estimate of the peak flux density for our candidates. The fluence of our candidates were then calculated using $\Phi_{\nu}=S_{\nu}W$, where $W$ is the pulse width measured at $50\%$ of the peak for a top-hat pulse \citep{Petroff..2019}. 

For LGRBs without any pulse candidates, we estimated the $6\sigma$ limit of their flux density and fluence. A maximum sample time of $t_{\rm{int}}=600$ ms was employed, corresponding to the largest downsampling factor used in our single-pulse search. Based on the radiometer equation, the flux density limit was then calculated. 
The fluence limits were then derived using a range of plausible pulse widths \citep[][equations 3 and 4]{Tian..2023}.
We assumed a rest-frame intrinsic pulse width of $W_{\rm{in}}=1$ ms, which is typical 
for FRBs \citep{CHIME..2021}. 
Taking into account the redshift \citep[assuming a typical LGRB of $z \sim 2$ or DM $ \sim 2000$ pc cm$^{-3}$;][]{Lan..2021}, dispersive smearing \citep[][equation 5]{Tian..2023},  
and pulse scattering from Galactic \citep{Bhat..2004,Price..2021} and host contributions at 185\,MHz \citep[][equation 29]{Xu..2016},
we derived the range of pulse width and the corresponding fluence constraints for all LGRBs without pulse candidates.

\section{Results} \label{sec:res}

Our single-pulse search results are summarized in Table \ref{tab:3}. The search results for reversed time series (see Section \ref{sec:search}) are also shown in Table \ref{tab:3}. To compare with the detection rate above $6\sigma$, we derive the expected number of false alarms above $6\sigma$ assuming Gaussian noise and independent trials \cite[see Section 4.1 from][]{Bannister..2012}. As shown in Table \ref{tab:3}, we find a similar number between expected false alarms and detected events above $6\sigma$, indicating that most candidate detections are likely noise \citep{Tian..2022pasa}. Based on the ``friends-of-friends'' algorithm, we find one candidate pulse from GRB $230123$A and one candidate pulse from GRB $230216$A. The relevant data for these two candidates are presented in Table \ref{tab:4}, which lists the pulse width, peak flux density, DM and corresponding redshift, pulse arrival time, and fluence. We note that for GRB $230123$A, a negative DM pulse search also yields a ``candidate'' from a cluster of detections above $6\sigma$ based on the {\sc rrattrap}. This unphysical candidate may be randomly generated from the noise fluctuation and reduces our confidence in the $1$ statistical positive detection. 

\begin{figure}[htbp]
	\centering
	\subfloat{
		\includegraphics[width=0.5\linewidth]{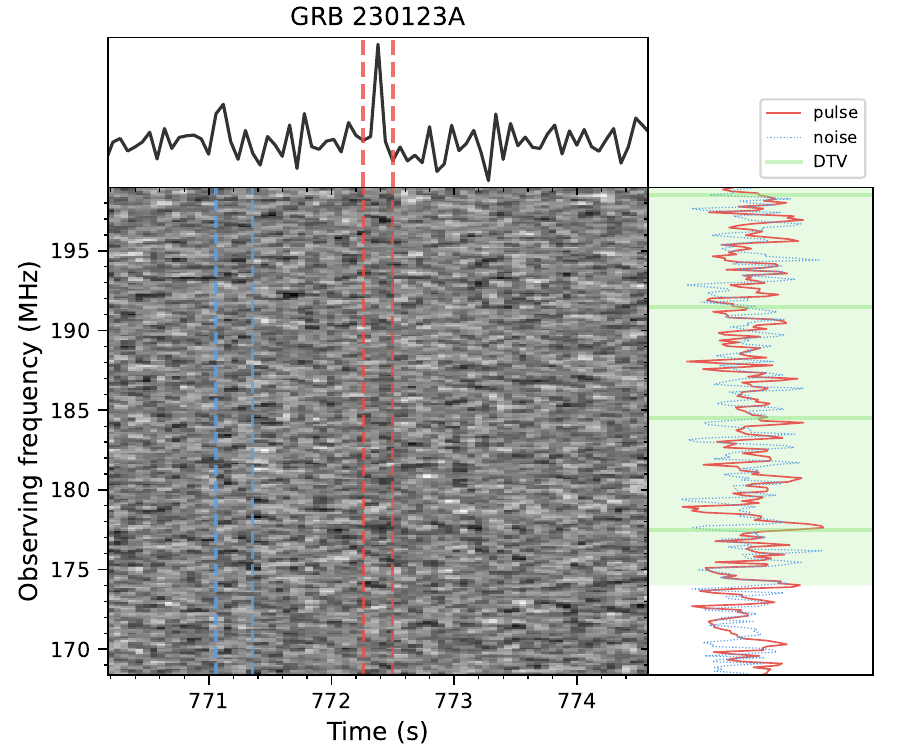}
	}
	\subfloat{
		\includegraphics[width=0.5\linewidth]{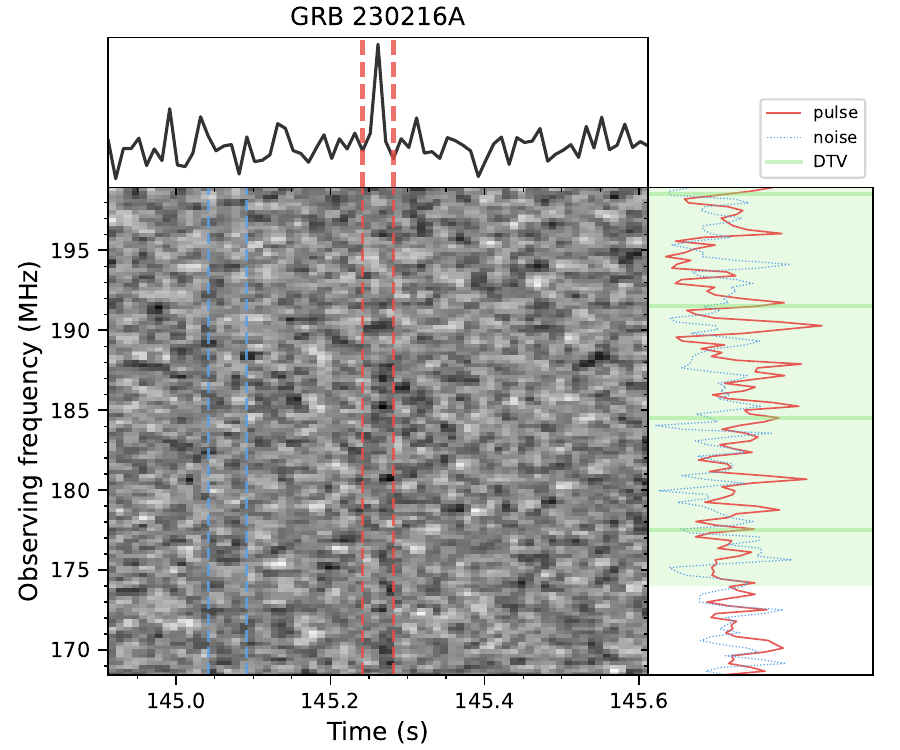}
	}	
	\caption{The pulse structure and the dynamic spectrum of our candidates. The left panel shows the dedispersed data (DM $=1404.12$ pc cm$^{-3}$) from GRB $230123$A with a time resolution of $60$ ms and a frequency resolution of $0.24$ MHz. The right panel shows the dedispersed data (DM $=409.89$ pc cm$^{-3}$) from GRB $230216$A with a time resolution of $10$ ms and a frequency resolution of $0.24$ MHz. The dynamic spectrum is smoothed with a uniform filter (see Section \ref{sec:res}). Displayed on the top of each panel is the frequency-averaged pulse profile. On the right side of each panel, we show the time-averaged spectrum for the pulse candidate (red solid line) and the background noise (blue dotted line). The corresponding regions are marked between two vertical dashed lines with the same colour in the dynamic spectrum. The background noise regions begin $20$ time bins before the pulse signal and span $5$ time bins in each panel. The DTV central frequencies are plotted over the spectrum with green horizontal lines and the DTV frequency bands are shown with light green regions. The time origin of these plots is set as the MWA observation start time. } 
	\label{fig:2}
\end{figure}

Figure \ref{fig:A1} in Appendix \hyperref[sec:appA]{A} shows our candidates from GRB $230123$A and GRB $230216$A on the DM-Time plot.
The pulse structure and the dynamic spectrum of the two candidates are presented in Figure \ref{fig:2}. The dynamic spectra are smoothed with a uniform filter from {\sc Scipy}. This smoothing process helps to diminish narrow-band structures and reduce instrumental noise \citep{Chen..2024}. Following the logic of \cite{Bannister..2012}, we calculate the confidence level of our candidates. We find that for both pulse candidates, the confidence level of them not being 
random noise fluctuations is $\sim 97\%$ ($2.2 \sigma$). In Figure \ref{fig:2}, we notice a faint linear structure in each panel that could represent a potential pulsed signal. These structures are also detected using the Hough transform technique that is widely used in identifying particular structures within an image (see Appendix \hyperref[sec:appB]{B} for details). However, comparing the spectrum of pulse candidates in Figure \ref{fig:2} with the background noise, we find they look very similar. Furthermore, from Figure \ref{fig:2}, we note that the pulse spectrum usually peaks at the DTV central frequencies and the DTV frequency bands almost cover the entire observing frequency band. \cite{Palaniswamy..2014} suggest that some low-level RFI may produce similar signals that pass the friends-of-friends test. Therefore, we find these candidates are likely RFI from DTV. 

We estimated the flux density limits, pulse width ranges and corresponding fluence limits for each GRB without candidates according to Section~\ref{sec:sen}, which are listed in Table~\ref{tab:5}.
There are several events with a significance above $6\sigma$ but did not pass the friends-of-friends test. These events are most likely RFIs.

\begin{table}[h!]
	\renewcommand{\thetable}{\arabic{table}}
	\centering
	\tabcolsep=6pt
	\renewcommand\arraystretch{1.2}
	\caption{Single-pulse search results and a comparison of high SNR events that emerge from the original and reversed time-series analysis described in Section\ref{sec:search}. } \label{tab:3}
	\begin{tabular}{lcccccc}
		\hline
		\hline
		GRB	name & Time series$^{a}$ & Search DM (pc cm$^{-3}$)	& Trial number$^{b}$ & 	Expected false alarms$^{c}$ & Detected events$^{c}$ & Candidates$^{d}$ \\
		\hline
		GRB 220427A & Original     & 100-5000	& 1.32e+10	& 26    & 26    & 0 \\
					& Reversed    &				&       	&     	& 40    & 0 \\
		\hline
		GRB 220518A & Original     & 100-5000	& 1.79e+10	& 35    & 52	& 0 \\
					& Reversed    &				&       	&     	& 74   	& 0 \\
		\hline
		GRB 221028A & Original     & 350-5000	& 2.92e+09	& 6     & 4		& 0 \\
					& Reversed    &				&       	&       & 3     & 0 \\
		\hline
		GRB 230123A & Original     & 100-5000	& 1.78e+10	& 35    & 63	& 1 \\
					& Reversed    &				&       	&       & 58    & 1 \\
		\hline
		GRB 230204A & Original     & 700-5000	& 1.30e+10	& 3     & 2     & 0 \\
					& Reversed    &				&       	&       & 0     & 0 \\
		\hline
		GRB 230216A & Original     & 150-5000	& 6.50e+09	& 13    & 26    & 1 \\
					& Reversed    &				&       	&       & 16    & 0 \\
		\hline
	\end{tabular}
	\begin{flushleft}
    	\tablenotetext{a}{The original time series search represents single-pulse search with positive DM and the reversed time series search represents single-pulse search with negative DM (see Section \ref{sec:search}).}
		\tablenotetext{b}{The trial number estimated from the number of time steps and DM trials.}
		\tablenotetext{c}{Counting events above $6\sigma$.}
  		\tablenotetext{d}{Events passing the friends-of-friends algorithm (see Section \ref{sec:search}).}
	\end{flushleft}
\end{table}

\begin{table}[h!]
	\renewcommand{\thetable}{\arabic{table}}
	\centering
	\tabcolsep=6pt
	\renewcommand\arraystretch{1.2}
	\caption{Pulse candidates. } \label{tab:4}
	\begin{tabular}{lccccccccc}
		\hline
		\hline
		GRB name	& Pulse SNR$^{a}$	& $T_{\rm{pulse}}^{b}$	& Pulse width$^{c}$  & Peak flux density$^{c}$ & DM$^{a}$       & $z_{\rm{pred}}^{d}$	& Fluence$^{e}$	&  $B^{f}$       & $P^{f}$ \\
					&					& (s)					& (ms)         & (Jy)              & (pc cm$^{-3}$) &			& (Jy ms)       & ($10^{15}$ G)  & (ms) \\		
		\hline
		GRB 230123A & 6.71	& 656.5 & $\sim$120	& 3.6$\pm$0.6	& 1404.12	& 1.36	& 431$\pm$74	& 6.7$\pm$1.1	& 8.5$\pm$0.6 \\
		GRB 230216A	& 6.82	& 155.7 & $\sim$20	& 10.5$\pm$1.9	& 409.89	& 0.18	& 211$\pm$37	& 18.1$\pm$5.4	& 35.3$\pm$7.0 \\
		\hline
	\end{tabular}	
	\begin{flushleft}
		\tablenotetext{a}{The SNR and DM of the pulse candidate obtained from the refined analysis (see Section \ref{sec:search}).}		
		\tablenotetext{b}{The dispersion corrected arrival time.}
		\tablenotetext{c}{The pulse widths and peak flux densities are both in the observer frame. }
		\tablenotetext{d}{The predicted redshift derived from the pulse DM (see Section \ref{sec:search}).}	
		\tablenotetext{e}{The fluence is calculated using $\Phi_{\nu}=S_{\nu}W$ (see Section \ref{sec:sen}).}	
		\tablenotetext{f}{Magnetar parameters using magnetar powered energy injection model (see Section \ref{sec:mod}). $B$ and $P$ represent the magnetic field strength and spin period, respectively. }	
	\end{flushleft}	
\end{table}

\begin{table}[h!]
	\renewcommand{\thetable}{\arabic{table}}
	\centering
	\tabcolsep=10pt
	\renewcommand\arraystretch{1.2}
	\caption{Flux density and fluence limits derived for LGRBs without candidates. } \label{tab:5}
	\begin{tabular}{lcccc}
		\hline
		\hline
		GRB name  & Galactic scattering time scale$^{a}$ (ms)	& Pulse width$^{b}$ (ms) & Flux density limit$^{c}$ (Jy) & Fluence limit$^{c}$ (Jy ms) \\
		\hline
		GRB 220427A & 2     & 26.1--666.6 & 3.6 & 94.8--2421.9 \\
		GRB 220518A & 0.1   & 26--664.6 & 1.5 & 39.6--1010.3 \\
		GRB 221028A & 38.4  & 46.4--702.9 & 1.4 & 62.7--949.4 \\
		GRB 230204A & 1451.3 & 1451.5--2115.5 & 3.7 & 5352.3--7800.4 \\
		\hline
	\end{tabular}
	\begin{flushleft}
		\tablenotetext{a}{Galactic scattering time derived from the NE2001 model.}	
		\tablenotetext{b}{The pulse range comes from the large uncertainty that lies in the scattering time scale (see Section \ref{sec:sen}).}	
		\tablenotetext{c}{$6 \sigma$ limit is derived here.}	
	\end{flushleft}
\end{table}

\section{Constraints on existing models} \label{sec:mod}

While RFI is the likely cause of these two candidates, we cannot completely rule out that they are real astrophysical signals. We therefore briefly investigated these signals in the context of their environments and theoretical constraints.

Before exploring the different emission models we might expect to produce prompt radio signals associated with a GRB, we first need to ensure the signal is not blocked by the surrounding environment. When the signal propagates to us, it may be blocked either by the dense shocked plasma in the GRB jet, which corresponds to the plasma frequency $\nu_{\rm{j}}$, or by the dense surrounding environment of LGRBs, which corresponds to the plasma frequency $\nu_{\rm{e}}$. 
The plasma frequency $\nu_{\rm{j}}$ can be calculated following \cite{Zhang..2014}, with the plasma density in the GRB jet given by $n^{\prime} \simeq 1.8 \times 10^{5} L_{x, 52} \Gamma_{2}^{-2} r_{17}^{-2} \mathrm{~cm}^{-3} $  (see Appendix \hyperref[sec:appC]{C} for details).The wind luminosity of the GRB, $L_{x}$, is assumed to be equal to the peak luminosity $L_{\rm{p}}$ at the prompt emission phase, as derived from the fitting results of the X-ray light curves for both candidates (see Appendix \hyperref[sec:appD]{D} for details). In our calculations of the jet plasma frequency, we adopt a typical bulk Lorentz factor of $\Gamma \sim 100$ and a shock radius of $r \sim 10^{17}$ cm \citep{Zhang..2018}. Using these parameters, we calculate the plasma frequency $\nu_{\rm{j}} \sim 161$ MHz for GRB $230123$A and $\nu_{\rm{j}} \sim 32$ MHz for GRB $230216$A in the observer frame. This means that the emission at $185$ MHz can escape the jet. For the plasma absorption from the surrounding environment, we use the effective neutral hydrogen column density $N_{\rm{H}}$ from the \emph{Swift}-XRT spectra\footnote{\url{https://www.swift.ac.uk/xrt\_spectra/}} and follow the procedure in \cite{Tian..2022mn} to calculate $\nu_{\rm{e}}$ (see Appendix \hyperref[sec:appC]{C} for details). We derive $\nu_{\rm{e}} \sim 12$ kHz for GRB $230123$A and $\nu_{\rm{e}} \sim 11$ kHz for GRB $230216$A in the ambient region. Therefore, the environment around both GRBs is transparent to our candidates.

We now explore the possible emission models from the candidate prompt signals from GRB $230123$A and GRB $230216$A. All models assume the GRB produces a magnetar remnant that drives the prompt emission. We used data products from the \emph{Swift} BAT-XRT unabsorbed flux light curve\footnote{\url{https://www.swift.ac.uk/burst\_analyser/}} \citep{Evans..2010}
to search for evidence of a magnetar remnant from both GRBs, which is usually observed as a plateau phase \citep{Zhang..2006,Rowlinson..2010,Rowlinson..2013}. Following methods outlined in \cite{Tian..2022mn}, we modeled the light curves of our candidates which both show plateaus (see Appendix \hyperref[sec:appD]{D}). Assuming these plateaus are from a magnetar central engine, we obtained the corresponding magnetic field $B$ and spin period $P$ for each candidate (see Table \ref{tab:4}). Since both candidates occurred several hundred seconds after their prompt emission, they are unlikely to be explained by the GRB jet-ISM model proposed by \cite{Usov..2000}. Additionally, both candidates show X-ray light curves with external plateau, indicating the formation of a stable magnetar. This rules out the possibility that the pulse candidates resulted from a magnetar collapsing into a black hole, as suggested by \cite{Falcke..2014}. On the other hand, the plateau phases observed in both candidates align with the predictions of the pulsar-like model proposed by \cite{Totani..2013}. In this model, the newly born millisecond magnetar generates coherent radio emission, and the spin-down luminosity of the magnetar can be calculated based on the magnetic field and spin period, which we derived from the light curve fitting (see Appendix \hyperref[sec:appD]{D}). Following their prescription, we constrained the radio emission efficiency for both GRBs to $\epsilon_{\rm{r}}\sim10^{-3}$ (see Figure \ref{fig:E1} in Appendix \hyperref[sec:appE]{E}) which is within the broad range of known pulsar emission efficiency from $10^{-7}$ to $1$ \citep{Szary..2014}. 


\section{Discussion} \label{sec:discuss}

\subsection{Challenge from RFI}

Narrow-band impulsive RFI poses a substantial challenge in accurately identifying signals, often resulting in misleading false detections. Since the DTV frequency bands almost cover our entire observing frequency band, it is most likely that our candidates are from narrow-band impulsive RFI related to digital TV. To mitigate this issue, we propose elevating the significance threshold for candidate detection in future analyses, ensuring that only the most robust signals are classified as authentic. 
Another option is to change the frequency range to a more RFI-quiet part of the MWA band, which also needs to be balanced against the loss of sensitivity due to scattering that causes pulse broadening. While the 30.72\,MHz bandwidth centred at 120\,MHz is much clearer of RFI \citep{Offringa..2015}, the pulse broadening is a factor of 5 greater than at 185\,MHz \citep{Bhat..2004}, attenuating the peak flux by the same amount and therefore reducing our sensitivity to pulsed signals.
A central frequency of 150\,MHz is a potential compromise that trades the broad DTV bands with brighter yet narrow-band RFI, combined with only a factor of 2 decrease in sensitivity due to scattering.
Lastly, conducting tests with off-source beams could be very useful in distinguishing these candidates as genuine.

\subsection{Comparison with other studies }

Assuming our detections are real, we can compare our results with previous studies. The first search for radio pulses from GRBs using the MWA was performed with imaging domain analysis \citep{Kaplan..2015,Anderson..2021pasa,Tian..2022pasa}. \cite{Kaplan..2015} conducted the first deep search for prompt radio emission from SGRB $150424$A, obtaining a flux density upper limit of $3$ Jy ($3 \sigma$) over $4$-second timescales at $80-133$ MHz, which is largely consistent with our results. Based on this flux limit, \cite{Rowlinson..2019} constrained the radio emission efficiency of SGRB $150424$A under the pulsar-like emission model to $\epsilon_{\rm{r}} \lesssim 10^{-2}$, which aligns with our finding of $\epsilon_{\rm{r}} \sim 10^{-3}$. Later, \cite{Anderson..2021pasa} presented the first GRB trigger (SGRB $180805$A) with the upgraded MWA rapid-response system at $170-200$ MHz, deriving an upper flux density limit of $270-630$ mJy ($3 \sigma$) over $5$-second timescales, comparable to our results. Similarly, \cite{Tian..2022pasa} conducted a low-frequency ($170-200$ MHz) search for coherent radio emission associated from nine SGRBs using the MWA. They found upper limits on flux density of a few Jy on $5$-second timescales. The radio emission efficiency for the pulsar-like emission model was constrained to $\epsilon_{\rm{r}} \lesssim 10^{-3}$, similar to our study. In a subsequent study, \cite{Tian..2022mn} reported the first GRB trigger (LGRB $210419$A) using the MWA VCS mode. They found an upper fluence limit of $77-224$ Jy ms for radio emission from GRB $210419$A, which is consistent with the limits derived in this study.

Several other attempts have been made to detect radio pulses from GRBs using LOFAR. For instance, \cite{Rowlinson..2019b} reported flux density limits of $1.7$ mJy ($3 \sigma$) over $2$ hours for LGRB $180706$A at $144$ MHz. \cite{Rowlinson..2021} searched for coherent radio emission following SGRB $181123$B, obtaining a $3 \sigma$ limit of $153$ mJy at $144$ MHz over a $136$-second integration time. Similarly, \cite{Hennessy..2023} searched for prompt radio emission associated with high-energy flares in LGRB $210112A$, deriving upper limits of $42$ mJy over $320$-second snapshot images and $87$ mJy over $60$-second images. These limits are lower than those derived in our study and indicate a radio emission efficiency for the pulsar-like model of less than $10^{-4}$, much smaller than our result of $\epsilon_{\rm{r}} \sim 10^{-3}$. In a recent study, \cite{rowlinson24} reported the detection of a candidate coherent radio flash from SGRB $201006$A. The signal was observed $76.6$ minutes after the burst with an SNR of $5.6 \sigma$ at $144$ MHz. The observed peak flux density was $49 \pm 27$ mJy in a $5$-second image, much fainter than the pulse candidates identified in our study. If the radio emission originates from persistent radiation of a newly born magnetar, the radio emission efficiency is constrained to $\epsilon_{\rm{r}} \sim 10^{-5}$, two orders of magnitude smaller than our estimate. 

It is worth noting that \cite{Bannister..2012} claimed to detected radio pulses from GRBs with significance above $6 \sigma$ at $1.4$ GHz using a single $12$ m dish antenna at the Parkes radio observatory. Assuming a spectral index of $-1.0$ \citep[based on the model described by][]{Totani..2013}, we estimate flux densities of $\sim0.5$ Jy and $\sim1.4$ Jy at $1.4$ GHz for GRB $230123$A and GRB $230216$A, respectively. These pulses would be too weak to be detected in the survey performed by \cite{Bannister..2012}. In contrast, the $7$ Jy pulse from GRB $101011$A detected by \cite{Bannister..2012} would be bright enough for us to detect under the spectral index assumption of $-1.0$.
	
Finally, \cite{Curtin..2023} used the Canadian Hydrogen Intensity Mapping Experiment (CHIME) Fast Radio Burst (CHIME/FRB) Project to explore potential coincidences between FRBs and GRBs. Although no statistically significant coincident pairs were found, they constrained radio flux density from LGRBs to be below a few kJy at $400-800$ MHz. Our pulse candidates, with flux densities of a few Jy at $400-800$ MHz (considering a spectral index of $-1.0$), fall well below their flux constraints.

\subsection{Unconventional model of LGRBs}

It is generally accepted that LGRBs should be produced by massive stellar collapse. Therefore, the radio pulse from LGRBs could mainly occur after the jet launch and during or at the end time of the plateau phase \citep{Rowlinson..2019}. It may also occur years after the LGRBs, but with weaker brightness \citep{Metzger..2017}. However, recent observations have shown that LGRBs could be generated from compact binary mergers \citep{Rastinejad..2022,Yang..2022,Levan..2024,Yang..2024}. In such a scenario, a pulse may also be emitted a few microseconds prior to the jet launch, namely a short-duration radio precursor \citep{Lipunov..1996,Metzger..2016}. The coherent radio emission could be released due to the dissipation of gravitational potential, electromagnetic energy, and rotational energy  \citep{Moortgat..2003,Pshirkov..2010,Lyutikov..2013,Mingarelli..2015,Metzger..2016,Cooper..2023}. The low-frequency rapid-response observation of the MWA actually provides us with a great opportunity to find such a signal. Assuming the delay between the GRB 
and MWA observation start time is $\lesssim60$\,s,
the dispersion of radio pulse at $185$ MHz will enable us to probe events
at DM$\gtrsim500$ pc cm$^{-3}$ or $z\gtrsim0.3$.

\section{Conclusion and future prospects} \label{sec:conclu}

The new generation of radio telescopes has greatly improved our ability to search for 
radio pulses associated with GRBs. In this study, we searched for radio pulse from six LGRBs at $185$ MHz with the MWA. Two plausible pulse candidates are detected with significance above $6\sigma$. The pulse from GRB $230216$A is more reliable than that from GRB $230123$A based on our reliability analysis. If the pulse comes from the pulsar-like emission of a newly born magnetar, the observed flux density constrains the radio efficiency as $\epsilon_{\rm{r}} \sim 10^{-3}$. This value is within the known pulsar emission efficiency range ($10^{-7}-1$). On the other hand, we could not rule out RFI from DTV as the origin of these pulses.

In the future, with a larger sample of LGRBs, we could expect to identify 
radio pulses with much higher SNR with the MWA. 
As demonstrated in Section \ref{sec:mod}, if the radio pulses can safely escape the local environment of LGRBs, we are likely to find more associated pulses with LGRBs than with SGRBs, given that LGRBs are more common. Additionally, redshift information will be a crucial asset in pulse searching as it allows us to 
narrow the DM searching range and help constrain the physical parameters of GRBs. Additionally, off-source beams could be generated to provide robust tests against RFI.

The SKA's low-frequency component (SKA-Low), which covers the frequency range of 50-350 MHz, is expected to provide significant sensitivity. Using the SKA sensitivity calculators\footnote{https://sensitivity-calculator.skao.int/low} \citep{Sokolowski..2022}, we find that with an integration time of $10$ ms at $200$\,MHz with a 300\,MHz bandwidth, the weighted continuum sensitivity is approximately $7$ mJy/beam (using a Briggs weighting scheme with robust set to 0). 
Assuming a pulse detection threshold of $7\sigma$, we will be capable of constraining pulses predicted by \citet{Totani..2013} down to a radio emission efficiency of $10^{-5}$ (as shown in Figure \ref{fig:E1}), and thus covering that of typical pulsars ($\epsilon_{\rm{r}} \approx 10^{-4}$).

\acknowledgments
We would like to thank the anonymous referee for helpful suggestions that lead to an overall improvement of this study. 
We also thank Apurba Bera and Xiang-Han Cui for valuable discussion. 

This scientific work uses data obtained from Inyarrimanha Ilgari Bundara / the Murchison Radio-astronomy Observatory. We acknowledge the Wajarri Yamaji People as the Traditional Owners and native title holders of the Observatory site. Establishment of CSIRO's Murchison Radio-astronomy Observatory is an initiative of the Australian Government, with support from the Government of Western Australia and the Science and Industry Endowment Fund. Support for the operation of the MWA is provided by the Australian Government (NCRIS), under a contract to Curtin University administered by Astronomy Australia Limited. This work was supported by resources provided by the Pawsey Supercomputing Research Centre with funding from the Australian Government and the Government of Western Australia.

This work also made use of data supplied by the UK \emph{Swift} Science Data Centre at the University of Leicester and the \emph{Swift} satellite. \emph{Swift}, launched in $2004$ November, is a NASA mission in partnership with the Italian Space Agency and the UK Space Agency. \emph{Swift} is managed by NASA Goddard. Penn State University controls science and flight operations from the Mission Operations Center in University Park, Pennsylvania. Los Alamos National Laboratory provides gamma-ray imaging analysis.

GEA is the recipient of an Australian Research Council Discovery Early Career Researcher Award (project number DE180100346).

F. X. acknowledges the great help from Xue-Feng Wu and Song-Bo Zhang to visit the International Centre for Radio Astronomy Research at Curtin University. F. X. also acknowledges the China Scholarship Program to conduct research at the International Centre for Radio Astronomy Research at Curtin University. 

Y.F.H. acknowledges the support from the Xinjiang Tianchi Program.

AR acknowledges funding from the NWO Aspasia grant (number: 015.016.033).

This study is supported 
by National Key R\&D Program of China (2021YFA0718500),
by National SKA Program of China No. 2020SKA0120300,
and by the National Natural Science Foundation of China (Grant Nos. 12233002).

The following software and packages were used to support this
work: 
{\sc MWA TRIGGER} \citep{Hancock..2019},
\footnote{\url{https://github.com/MWATelescope/mwa\_trigger/}} 
{\sc COMET} \citep{Swinbank..2014}, 
{\sc VOEVENT-PARSE} \citep{Staley..2016}, 
{\sc ASTROPY} \citep{Astropy..2022}, 
{\sc NUMPY} \citep{Harris..2020}, 
{\sc SCIPY} \citep{Virtanen..2020}, 
{\sc MATPLOTLIB} \citep{Hunter..2007},
{\sc PRESTO} \citep{Ransom..2001}. 
This research has made use of NASA’s Astrophysics Data System.

\nocite{*}
\bibliographystyle{aasjournal}
\bibliography{bibtex}

\appendix

\section{Candidates pass the friends-of-friends test} \label{sec:appA}

\begin{figure*}[htbp]
	\centering
	\includegraphics[width=0.8\linewidth]{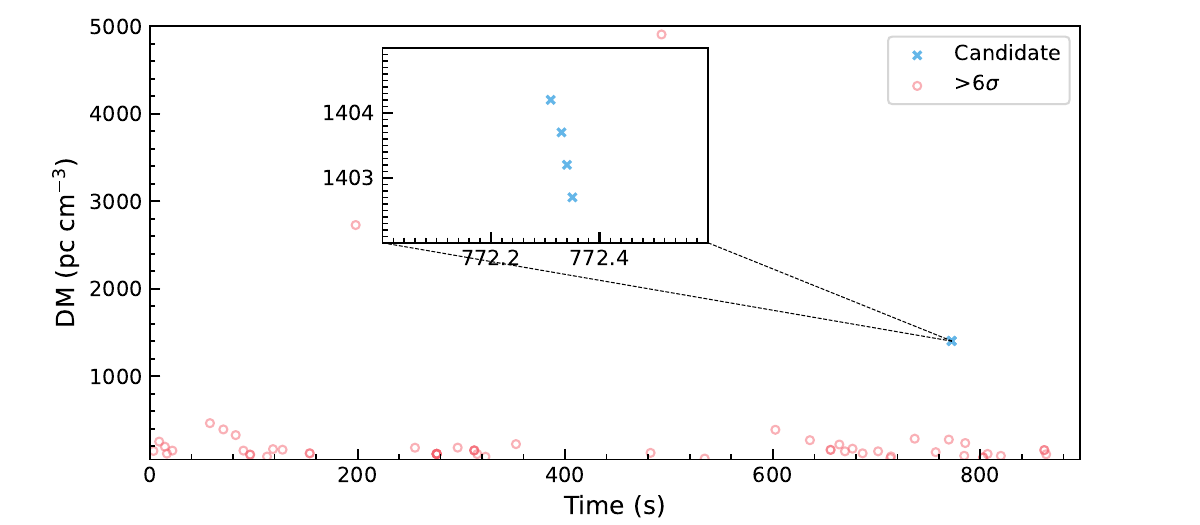}
	\includegraphics[width=0.8\linewidth]{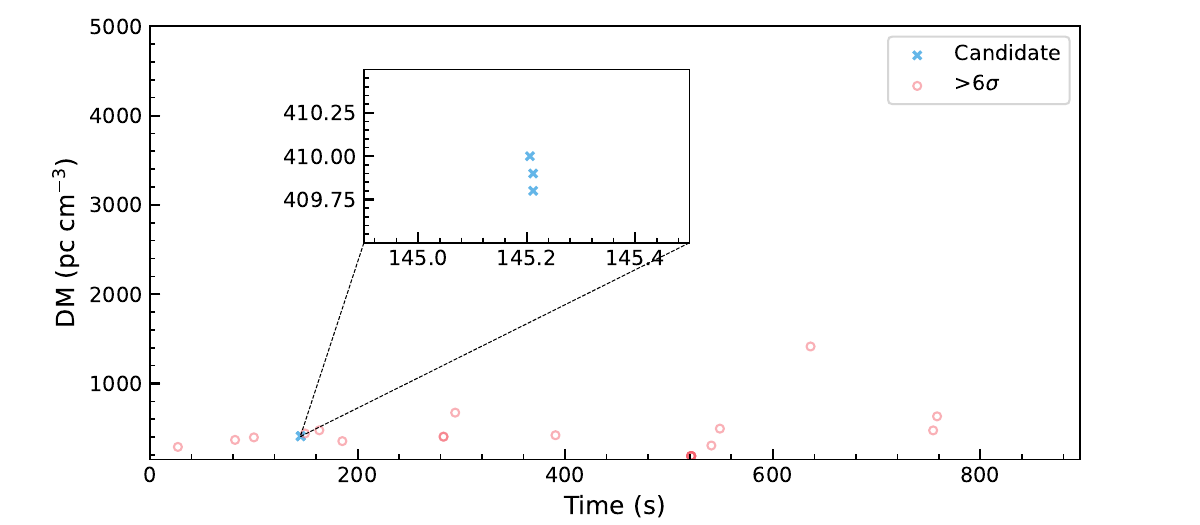}
	\caption{DM-Time plot for two GRBs with candidates. Detected events with significance above $6\sigma$ are marked on these plots. We classify a candidate (blue crosses) if it passes the ``friends-of-friends'' algorithm (see Section \ref{sec:search}). All other $> 6\sigma$ events are plotted with red circles. The time axis begins at the moment the telescope first targets the source. } 
	\label{fig:A1}
\end{figure*}

In Figure \ref{fig:A1}, we present our candidates from GRB $230123$A and GRB $230216$A on the DM-Time plot. We also plot other detected events that have significance above $6\sigma$ that did not pass the ``friends-of-friends'' test in this figure. 

\section{Identifying linear structure with Hough Transform} \label{sec:appB}

The Hough transform, introduced by Hough \citep{Hough..1962}, is a coordinate transformation commonly used to identify particular structures inside an image. It converts points of interest from the $x$-$y$ image space into a line in the $\theta$-$\rho$ parameter space using the following parameterization

\begin{equation}
	\rho = x \cos (\theta) + y \sin (\theta).
\end{equation}

In this work, we focus on identifying faint linear structures from original figures (Figure \ref{fig:2}), where points of interest are extracted by retaining pixels with gray values lower than $90$. This threshold is arbitrarily set to reduce noise and ensure sufficient features for the Hough transform. The Hough transform is implemented using the {\sc OpenCV} library. The resolution of the $\theta$ and $\rho$ parameters are set to $0.1^\circ$ and $0.1$, respectively. The accumulator thresholds are set as $160$ and $220$ for GRB $230123$A and GRB $230216$A, respectively. 
	
\begin{figure}[htbp]
	\centering
	\includegraphics[width=0.45\textwidth]{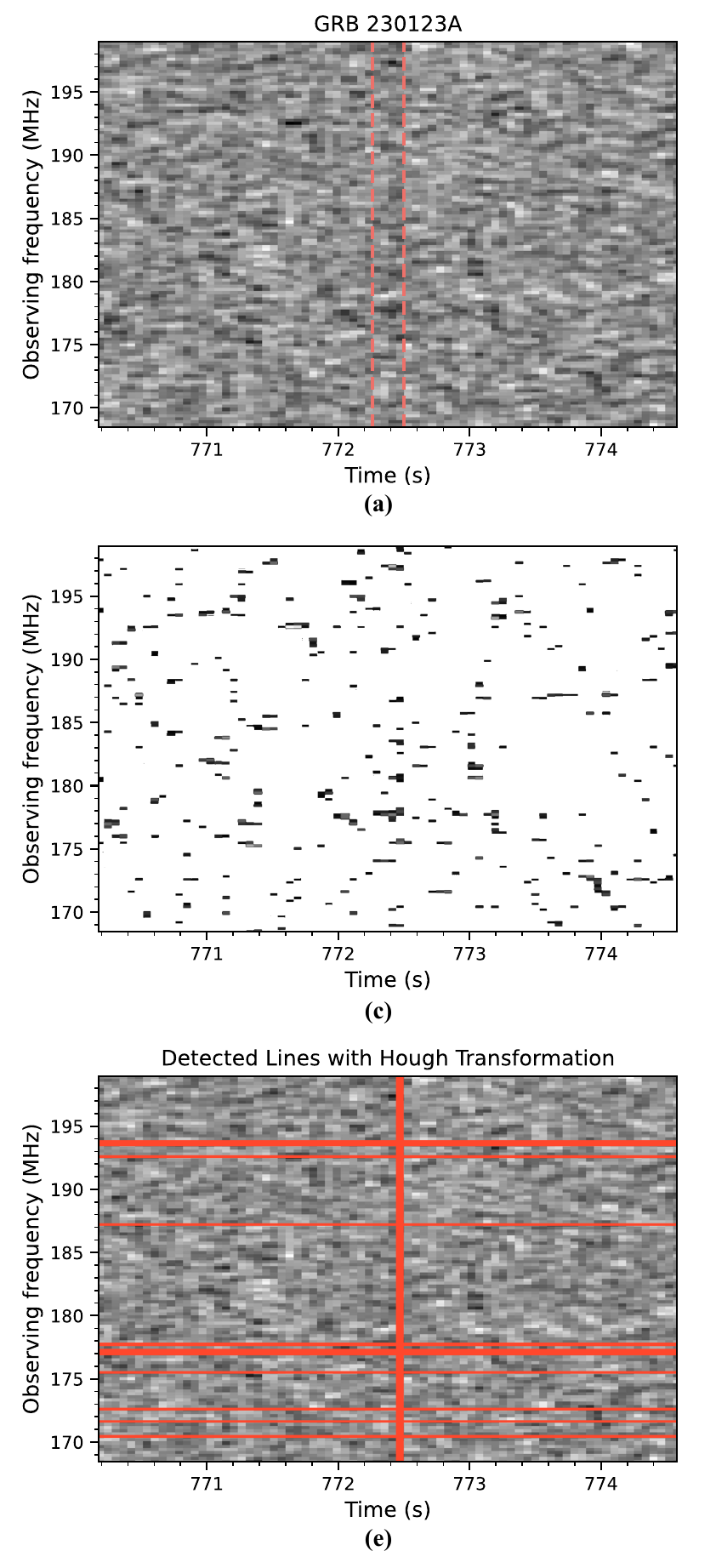}
	\includegraphics[width=0.45\textwidth]{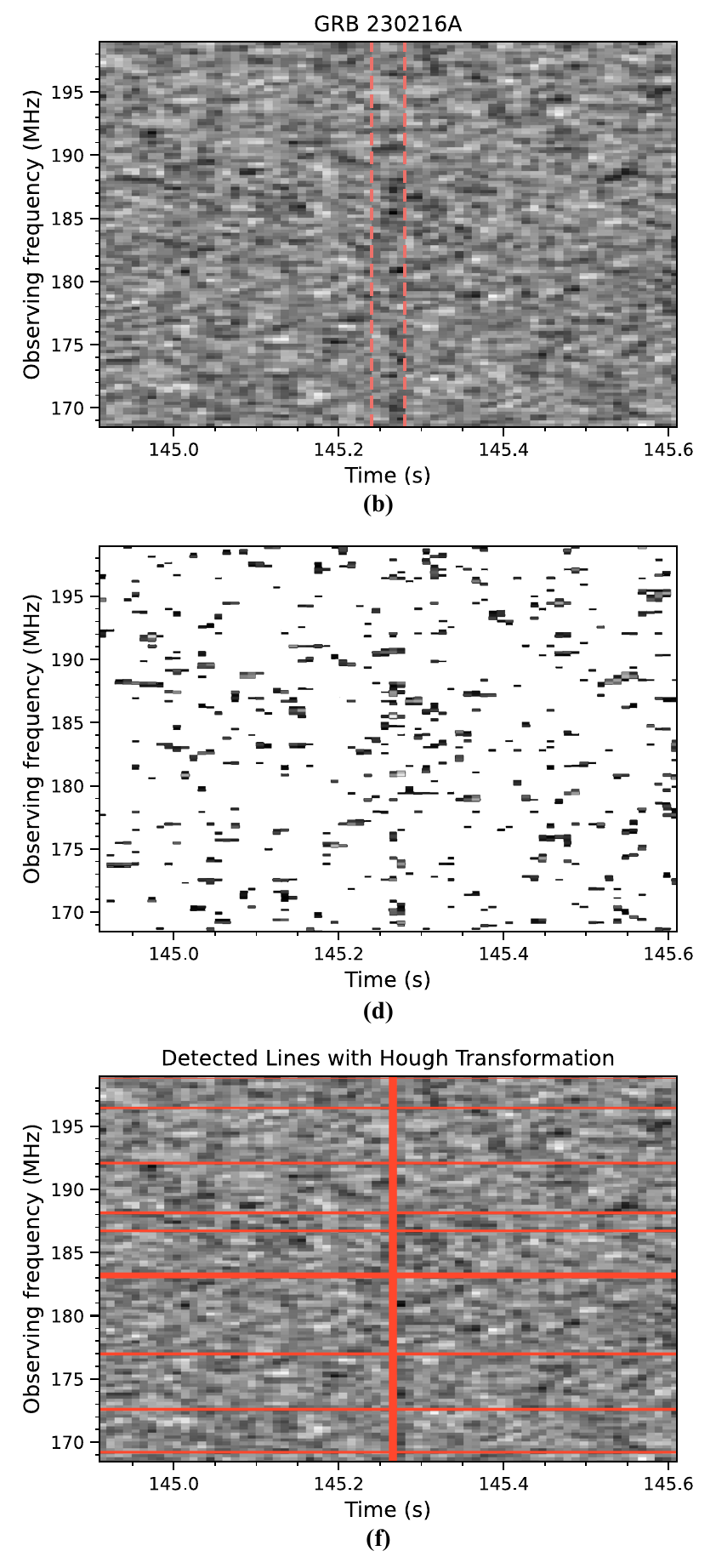}
	\caption{ Identifying of faint linear structures using the Hough transform. Panels (a) and (b) show the original images with possible pulse candidates, marked between two red vertical dashed lines. Panels (c) and (d) display the points of interest extracted from the origin images. Panels (e) and (f) show the final results, with red solid lines representing the identified lines. } 
	\label{fig:B1}
\end{figure}

Figure \ref{fig:B1} illustrates the results. Figures \ref{fig:B1}(a) and (b) display the original images (Figure \ref{fig:2}) with possible pulse candidates. Figures \ref{fig:B1}(c) and (d) present the extracted features, obtained after applying the gray value thresholding. The detected lines are highlighted as red solid lines in Figure \ref{fig:B1}(e) and (f). The Hough transform successfully figures out faint vertical lines, which are possible pulse signals. Additionally, several horizontal lines are also identified, which are attributed to noise present in the feature extraction images (Figure \ref{fig:B1}(c) and (d)).

\section{Calculation of plasma's oscillation frequency} \label{sec:appC}

An important issue for low-frequency radio emission searches is plasma absorption. The signal can be  blocked if the plasma's oscillation frequency, $\nu_{\rm{p}}$, exceeds the observation frequency. The plasma frequency can be calculated with 

\begin{equation}
	\nu_{\rm{p}}=\left( \frac{ne^{2}}{\pi m_{\rm{e}}} \right)^{1/2},
\end{equation}
where $e$ and $m_{\rm{e}}$ are the charge and mass of electrons respectively and $n$ is the electron number density in the plasma.

As proposed by \cite{Zhang..2014}, the GRB shock in front of the radio emission region may have the highest electron density to block the emission. Following \cite{Zhang..2014}, the plasma density in the jet comoving frame can be calculated as 

\begin{equation}
	n^{\prime} \simeq 1.8 \times 10^{5} L_{\rm{w}, 52} \Gamma_{2}^{-2} r_{17}^{-2} \mathrm{~cm}^{-3},
\end{equation}
where $L_{\rm{w}}$ is the wind luminosity of the GRB, $\Gamma$ is the bulk Lorentz factor, and $r$ is the shock radius. Therefore, in the observer frame, the plasma frequency $\nu_{\rm{j}}$ at the GRB jet shock front is

\begin{equation}
	\nu_{\rm{j}}=\frac{\Gamma}{(1+z)} \left( \frac{n^{\prime} e^{2}}{\pi m_{\rm{e}}} \right)^{1/2} \simeq 380~(1+z)^{-1} L_{\rm{w}, 52}^{1/2} r_{17}^{-1} \mathrm{~MHz}.
\end{equation}
To assess whether our candidates can escape the GRB shock front, we assume that the wind luminosity $L_{\rm{w}}$ is equal to the peak luminosity of the prompt emission phase. From the X-ray light curve fitting, we estimate $L_{\rm{w}}\sim10^{52}$ erg s$^{-1}$ for GRB $230123$A and $L_{\rm{w}}\sim10^{50}$ erg s$^{-1}$ for GRB $230216$A (see Appendix \hyperref[sec:appD]{D} for details). With the predicted redshift listed in Table \ref{tab:4} derived from the DM of pulse candidates and considering a typical shock radius of $r \sim 10^{17}$ cm \citep{Zhang..2018}, we obtain $\nu_{\rm{j}} \sim 161$ MHz for GRB $230123$A and $\nu_{\rm{j}} \sim 32$ MHz for GRB $230216$A in the observer frame. Therefore both candidate pulses can pass through the GRB shock front.

Another region that may block the radio emission is the surrounding environment of LGRBs as suggested by \cite{Tian..2022mn}. It is commonly accepted that LGRBs are located within star-forming regions in their host galaxies where the local density is higher than the normal diffuse ISM \citep{woosley93,Paczynski..1998,Galama..2001,Tam..2005}. The X-ray spectra of LGRBs provide valuable insights into the circumburst environment \citep{Lyman..2017,Heintz..2018}. Some LGRBs are found to have a large effective neutral hydrogen column density $N_{\rm{H}}$ based on X-ray spectral fitting \citep{Galama..2001,Prochaska..2007}. This has been suggested as evidence that LGRBs are associated with giant molecular clouds by \cite{Galama..2001}. The effective neutral hydrogen density of the giant molecular clouds can be estimated as 

\begin{equation}
	n_{\rm{H}}=500 \frac{N_{\rm{H}}}{10^{22.5} \mathrm{~cm}^{-2}} (\frac{R}{20 \mathrm{~pc}})^{-1}  \mathrm{~cm}^{-3},
\end{equation}
where $R$ is the radius of clouds \citep{Galama..2001}. Considering absorbed power-law fits to the \emph{Swift} XRT spectra\footnote{\url{https://www.swift.ac.uk/xrt\_spectra/}} \citep{Evans..2007,Evans..2009}, we have $N_{\rm{H}}=1.26 \times 10^{21}$ cm$^{-2}$ for GRB $230123$A and $N_{\rm{H}}=8.8 \times 10^{20}$ cm$^{-2}$ for GRB $230216$A. These values are derived under the assumption of zero redshift and Solar metallicity for the gas, due to the lack of detailed host information. Nevertheless, the obtained values are in agreement with the typical $N_{\rm{H}}$ value of LGRBs, which is $5^{+10.8}_{-3.4} \times 10^{21}$ cm$^{-2}$ \citep{Campana..2012}. Therefore, we consider these values to be reasonable estimates for the actual gas column densities. The cloud radius can be typically taken as $20$ pc. The plasma frequency of the surrounding environment of GRBs $\nu_{\rm{e}}$ in the observer frame can be calculated with

\begin{equation}
	\nu_{\rm{e}}=\frac{1}{(1+z)} \left( \frac{n e^{2}}{\pi m_{\rm{e}}} \right)^{1/2} = \frac{1}{(1+z)} \left( \frac{\chi n_{\rm{H}} e^{2}}{\pi m_{\rm{e}}} \right)^{1/2} \simeq 9~(1+z)^{-1} \chi^{1/2} n_{\rm{H}}^{1/2} \mathrm{~kHz},
\end{equation}
where $\chi$ is the ionization of the clouds. The ionizing photons from the GRB can ionize the IGM along the line-of-sight, potentially up to several hundreds of parsecs \citep{Ledoux..2009,Saccardi..2023}.
Therefore, we assume that the IGM along the line-of-sight is fully ionized, i.e., $\chi \lesssim 1$. In this way, we can calculate the plasma frequency of the surrounding environment for each GRB. For GRB $230123$A, we obtain $\nu_{\rm{e}} \sim 17$ kHz, while for GRB $230216$A, we derive $\nu_{\rm{e}} \sim 28$ kHz. Therefore the GRB ambient region is transparent for our pulse candidates.

\section{Fitting the X-ray Light curve and deriving the magnetar properties} \label{sec:appD}

The prompt emission model explored in Section \ref{sec:mod} assumes the formation of a magnetar remnant, which is one possible source of the plateau phase commonly observed in GRB X-ray light curves (e.g. \cite{Rowlinson..2010,Rowlinson..2013}). In the following, we describe the fitting process for the \emph{Swift} BAT and XRT light curves of GRB $230123$A and GRB $230216$A where the plateau phases are used to infer the properties of their possible magnetar remnants. Combining the \emph{Swift} BAT and XRT unabsorbed flux data \citep{Evans..2010}, we construct $0.3-10$ keV light curves for both candidates. To better distinguish the plateau in the afterglow phase observed by XRT, it is more convenient and straightforward to directly use the $0.3-10$ keV observation band of XRT. This approach minimizes errors that could arise from extrapolating beyond the observed energy range of the XRT data. The flux is then converted into luminosity with the predicted redshift listed in Table \ref{tab:4}. We take a flat $\Lambda$CDM model with $H_{0}=67.8$ km s$^{-1}$ Mpc$^{-1}$ and $\Omega_{\mathrm{m}}=0.308$ \citep{Planck..2016} to calculate the luminosity distance. Furthermore, we apply the k-correction to obtain the bolometric luminosity $L$ in the $[1/(1+z)-10000/(1+z)]$ keV energy range. To calculate the k-correction factor, we consider a simple power-law spectrum with the spectral index taken from data product of \emph{Swift} XRT spectra \citep{Evans..2007,Evans..2009}. 

The prompt emission and the flare can be fitted by a segmentation function

\begin{equation}
	L=\left\{\begin{array}{ll}
		L_{\rm{p}} e^{a(t-t_{\rm{p}})}, \ \  & t \leq t_{\rm{p}}, \\
		L_{\rm{p}}(\frac{t}{t_{\rm{p}}})^b, \ \ & t > t_{\rm{p}},
	\end{array}\right.
\end{equation}
where $t_{\rm{p}}$ and $L_{\rm{p}}$ represent the peak time and peak luminosity, respectively. The parameters $a$ and $b$ serve as the rise and decay parameters, and $t$ is the observation time. 

The plateau is modeled as the result of energy injection from a millisecond magnetar, given by

\begin{equation}
	L=L_{\rm{sd}}\left(1+\frac{t}{T_{\rm{sd}}}\right)^{-2},
\end{equation}
with the spin-down luminosity $L_{\rm{sd}} = \frac{B^2 R^6(2 \pi)^4}{6 c^3 P^4} = \frac{8\pi^4 B^2 R^6}{3 c^3 P^4}$ and the spin-down time scale $T_{\rm{sd}} = (1+z)\frac{3 c^3 I P^2}{B^2 R^6(2 \pi)^2} = \frac{3 (1+z) c^3 I P^2}{4\pi^2 B^2 R^6}$ \citep{Shapiro..1983,Zhang..2001}. Here $c$ is the speed of light, $B$ and $P$ are the magnetic field strength and the spin period of the millisecond magnetar, respectively. We adopt a magnetar radius of $R=10^{6}$ cm and a moment of inertia as $I=10^{45}$ g cm$^{2}$. The plateau luminosity is primarily determined by the magnetic field strength $B$ and the spin period $P$ of the magnetar remnant. 

\begin{figure}[htbp]
	\centering
	\includegraphics[width=0.48\textwidth]{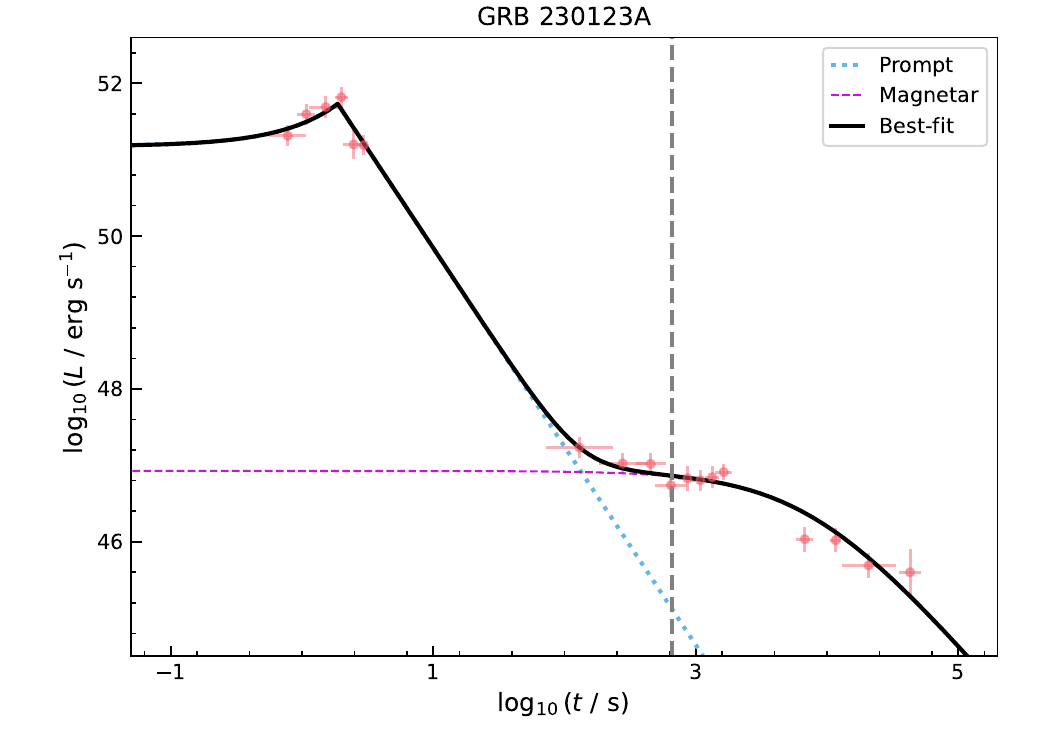}
	\includegraphics[width=0.48\textwidth]{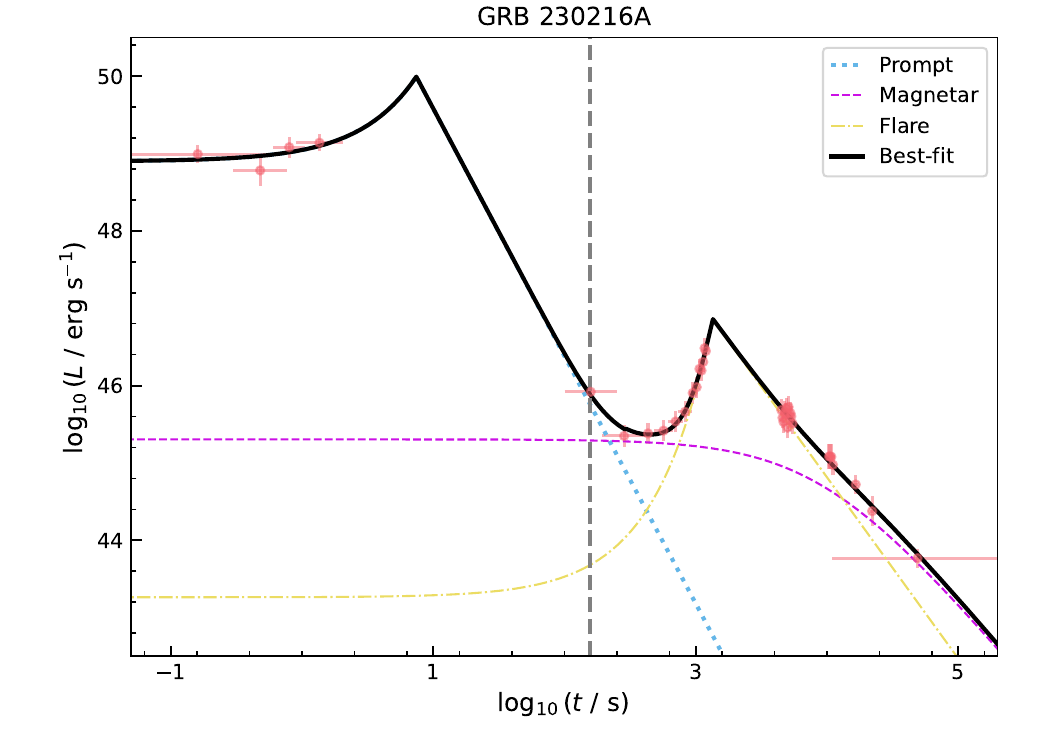}
	\caption{ The X-ray light curves of two GRBs with candidate radio pulses. The left panel corresponds to GRB $230123$A, and the right panel corresponds to GRB $230216$A. The red points represent the \emph{Swift} BAT and XRT data, with error bars indicating $1 \sigma$ uncertainty. The best-fit results are shown as black solid curves. The prompt emission component, the magnetar produced plateau component, and the flare component are marked with blue dotted line, purple dashed line, and yellow dot-dashed line, respectively. The emission time of the pulse candidates in the observer frame $T_{\rm{pulse}}$ are marked with gray dashed vertical lines. } 
	\label{fig:D1}
\end{figure}

We apply the above prescription to model the light curves of GRB $230123$A and GRB $230216$A as illustrated in Figure \ref{fig:D1}. Both candidates show plateaus lasting up to $\sim 10^{4}$ s. For GRB $230123$A, the light curve includes both prompt and plateau components (see the left panel). On the other hand, an additional flare component occurring between $1000$ and $3000$ s after the burst is required for GRB $230216$A (see the right panel of Figure \ref{fig:D1}). In Figure \ref{fig:D1}, we also mark the dispersion delay corrected time of our pulse candidates with gray dashed vertical lines. For GRB $230123$A, the pulse was emitted during the plateau phase, while for GRB $230216$A, the pulse occurred at the beginning of the plateau phase. Best-fit results with $1\sigma$ uncertainties for the magnetic field $B$ and the spin period $P$ of a magnetar central engine for both GRBs are listed in Table \ref{tab:4}. Our fitting results are consistent with typical values derived in previous studies \citep{Rowlinson..2013,Lu..2014,Li..2018,Rowlinson..2019,Rowlinson..2021,Tian..2022pasa}. 

\section{Constraints on the radio emission efficiency} \label{sec:appE}

\begin{figure}[htbp]
	\centering
	\includegraphics[width=0.95\textwidth]{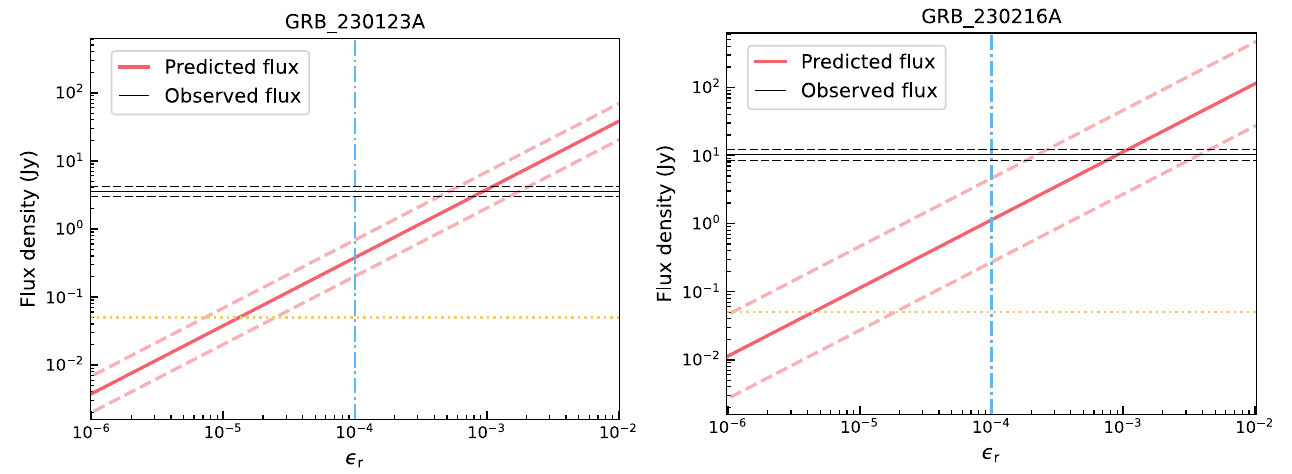}
	\caption{Constraints on the radio emission efficiency from the pulsar-like emission model (see Section \ref{sec:mod}). The red thick solid lines correspond to the predicted flux density. The black thin solid lines represent the observed flux density. The dashed lines represent the $1 \sigma$ confidence level. The blue vertical line represents the typical emission efficiency observed for known pulsars. The yellow dotted line represents the SKA-Low sensitivity to a $10$ ms pulse assuming a $7 \sigma$ threshold.
    } 
	\label{fig:E1}
\end{figure}

In Figure \ref{fig:E1}, we present the expected flux density as a function of $\epsilon_{\rm{r}}$ based on the model proposed by \cite{Totani..2013}. Compared with the observed flux density of our candidates, we obtain $\epsilon_{\rm{r}}=10^{-3.02\pm0.34}$ for GRB $230123$A and $\epsilon_{\rm{r}}=10^{-3.03\pm0.69}$ for GRB $230216$A. These results suggest higher values than the typical pulsar radio emission efficiency of $10^{-4}$ but still within the broad range of pulsar emission efficiency from $10^{-7}$ to $1$ \citep{Szary..2014}.

\end{document}